\documentclass[aps, prd, onecolumn, tightenlines, notitlepage, superscriptaddress, nofootinbib, preprintnumbers, floatfix,showkeys]{revtex4-2}

\usepackage{amstext}
\usepackage{amssymb}
\usepackage{amsmath}
\usepackage{graphicx}
\usepackage{url}
\usepackage{color}
\usepackage[utf8]{inputenc}
\pdfoutput=1
\usepackage{textcomp}

\usepackage{epsfig,amsfonts,mathrsfs,amsmath,amssymb,graphicx,color,slashed,multirow}
\usepackage{amsmath,latexsym,amssymb,graphicx,slashed,color,enumerate,url,cancel,gensymb}
\usepackage{textcomp}

\usepackage[x11names]{xcolor}
\usepackage[colorlinks]{hyperref}
\usepackage[normalem]{ulem}

\definecolor{nicered}{rgb}{0.7,0.1,0.1}
\definecolor{nicegreen}{rgb}{0.1,0.5,0.1}

\AtBeginDocument{\hypersetup{citecolor=Green3,linkcolor=Red1,urlcolor=blue}}

\begin{document}


\title{{\Large Neutrino CPT violation in the solar sector}}

\author{G. Barenboim}\email{gabriela.barenboim@uv.es}
\affiliation{Instituto de Física Corpuscular, CSIC-Universitat de València, C/Catedrático José Beltrán 2, Paterna 46980, Spain}
\affiliation{Departament de Física Teòrica, Universitat de València,
C/ Dr. Moliner 50, Burjassot 46100, Spain}

\author{P. Mart\'{i}nez-Mirav\'e}\email{pamarmi@ific.uv.es}
\affiliation{Instituto de Física Corpuscular, CSIC-Universitat de València, C/Catedrático José Beltrán 2, Paterna 46980, Spain}
\affiliation{Departament de Física Teòrica, Universitat de València,
C/ Dr. Moliner 50, Burjassot 46100, Spain}

\author{C.A. Ternes}\email{ternes@to.infn.it}
\affiliation{Istituto Nazionale di Fisica Nucleare (INFN), Sezione di Torino, Via P. Giuria 1, I--10125 Torino, Italy}
\affiliation{Dipartimento di Fisica, Universit\`a di Torino, via P. Giuria 1, I–10125 Torino, Italy}

\author{M. Tórtola}\email{mariam@ific.uv.es}
\affiliation{Instituto de Física Corpuscular, CSIC-Universitat de València, C/Catedrático José Beltrán 2, Paterna 46980, Spain}
\affiliation{Departament de Física Teòrica, Universitat de València,
C/ Dr. Moliner 50, Burjassot 46100, Spain}

\begin{abstract}
In this paper we  place new bounds on CPT violation in the solar neutrino sector analyzing the results from solar experiments and KamLAND.
We also discuss the sensitivity 
of the next-generation experiments DUNE and Hyper-Kamiokande,   
which will provide accurate 
measurements of the solar neutrino oscillation parameters. 
The joint analysis of both experiments will further improve the precision due to cancellations in the systematic uncertainties regarding the solar neutrino flux. In combination with the next-generation reactor experiment JUNO, the bound on CPT violation in the solar sector could be improved by one order of magnitude in comparison with current constraints.
The distinguishability among CPT-violating neutrino oscillations and neutrino non-standard interactions in the solar sector is also addressed.
\end{abstract}


\maketitle

\section{Introduction}
Symmetries are a key ingredient of any area of science. They are present in chemistry and biology, but it is indeed in particle physics where they play a fundamental role. There, the structure of non-elementary particles themselves and their interactions is deduced from symmetries and invariance properties.
The very existence of particle masses follows the same lines.
Among all the symmetries present in particle physics, CPT invariance played a defining role.
It surely is one of Nature's most essential symmetries  and its invariance has been used as a guiding tool to construct models. This is why its experimental validation is so crucial and the reason behind so many experimental efforts to test it~\cite{Widmann:2021krf}.

In particle physics, field theory is the mathematical tool to construct models. Through the field theory lens, perhaps  one of the most intriguing properties of free antiparticles, courtesy of CPT invariance, is that they may be mathematically regarded as if they were  simply particles with the same mass and  opposite charge (as compared with their counterparts), but moving backwards in time and space.
Clearly, CPT invariance is embedded in the founding pillars of our model since its construction.
Shortly, it states that the three independent operations of charge conjugation (C),  parity (P), and time reversal (T), if simultaneously performed, would not modify any measurable property of the system.
Furthermore, the CPT theorem \cite{Streater:1989vi} guarantees that any local, relativistic quantum field theory that preserves Lorentz invariance automatically conserves CPT. And precisely because of its  anti-unitary nature, the CPT operator connects the S-matrix of a process to the inverse process's S-matrix, where particles are replaced by their antiparticles and all the spin components are reversed. Needless to say, this does not imply that two CPT conjugated processes will have the same probability. A particle will not decay at the same rate as its antiparticle to the same final states. Instead, the sum of all partial decay rates, the  lifetime of this particle, would be the same as that of its antiparticle. 

Precisely because it is so deeply buried in field theory, its evaluation using elementary particles, and not composites, becomes essential, and the best system where this can be done is in neutrino experiments. Certainly, a potential discovery of CPT violation would imply that at least one of its key axioms like Lorentz invariance, interaction locality, or unitarity must be rejected.
Neutrinos are not only the ideal system for CPT to be tested, they offer also the best opportunity to do it from an experimental as well as a
theoretical point of view.  
The origin of neutrino mass and its smallness is not satisfactorily addressed by the  Standard Model and its incorporation generally comes along with new particles, new interactions and new scales. 
Some plausible explanations invoke the existence of a very high scale, not far from where we expect gravity to be non-local, and where new physical laws can appear or the old ones get modified to incorporate Lorentz and CPT violation. Then, neutrinos can be thought of as our window to such high scales and our chance to test fundamental symmetries to an unprecedented level.

In the standard picture of the three-neutrino paradigm, the neutrino oscillation parameters are fairly well measured, see Refs.~\cite{deSalas:2020pgw,Capozzi:2021fjo,Esteban:2020cvm}. However, it is well known that new physics might affect these measurements. In this paper, we assume a non-local field theory~\cite{Barenboim:2002tz} where different mass terms for neutrinos and antineutrinos can be present, allowing for different mixing parameters. Hence, we consider that flavor oscillations of neutrinos are described by a set of oscillation parameters, $\Delta m_{ji}^2$, $\theta_{ij}$ and $\delta$, whereas antineutrinos oscillate with $\Delta \overline{m}_{ji}^2$, $\overline{\theta}_{ij}$ and $\overline{\delta}$. If CPT is conserved, the parameters must coincide (except for CP-violating effects). This formalism has been used to compute bounds on the differences between neutrino and antineutrino oscillation parameters~\cite{Barenboim:2017ewj,Tortola:2020ncu}. These works focused on analyzing data from accelerator and reactor experiments~\cite{Tortola:2020ncu}, while also discussing future sensitivities at DUNE~\cite{Barenboim:2017ewj}. Sensitivity studies for other experiments have also been performed in Refs.~\cite{deGouvea:2017yvn,Majhi:2021api,Ngoc:2022uhg}. In this paper, we extend the discussion of Refs.~\cite{Barenboim:2017ewj,Tortola:2020ncu} to the solar sector. The solar neutrino parameters have been measured using solar neutrinos observed at many experiments and reactor antineutrinos using the KamLAND detector. While the measurement of the mixing  angles $\theta_{12}$ and  $\overline{\theta}_{12}$ showed good agreement\footnote{Note that KamLAND is basically octant-blind and that if CPT was violated such that the neutrino parameter was $\sin^2\theta_{12} \approx 0.3$ and its antineutrino counterpart $\sin^2\overline{\theta}_{12} \approx 0.7$, we would have no possibility to observe it.}, there was a small tension between the measurement of $\Delta m_{21}^2$ and $\Delta \overline{m}_{21}^2$. This tension has been at the 2$\sigma$ level for many years, while with the latest preliminary Super-Kamiokande solar data it is reduced to the $\sim1.4\sigma$ level~\cite{yusuke_koshio_2022_6695966}.

In this paper, we discuss how this tension between the neutrino and antineutrino solar mass splitting can evolve in the future. First, we calculate a new bound on the differences between neutrino and antineutrino oscillation parameters using the most recent data in Section~\ref{sec:bounds}. We briefly summarize the simulation details of the future experiments JUNO~\cite{JUNO:2015zny}, Hyper-Kamiokande~\cite{Hyper-Kamiokande:2018ofw} and DUNE~\cite{DUNE:2020lwj,DUNE:2020ypp,DUNE:2020mra,DUNE:2020txw} in Section~\ref{sec:sim}, while in Section~\ref{sec:future} we discuss the bound that can be obtained from the combined analysis of reactor and solar experiments. In Section~\ref{sec:cpt_measure} we estimate the significance of the tension if the best fit values of the measurements do not change. Next, in Section~\ref{sec:cpt-nsi} we discuss the possibility of interpreting a CPT-violating signal in terms of neutrino nonstandard interactions (NSI) in matter. Finally,  we conclude in Section~\ref{sec:conc}.

\section{Updated bounds on CPT violation in the solar sector}
\label{sec:bounds}

As a first step, we update the bounds on CPT-violating neutrino and antineutrino oscillation parameters in the solar sector. For a given neutrino oscillation parameter $x$ and its antineutrino counterpart $\overline{x}$, the limit on CPT violation is obtained by evaluating 
\begin{equation}
    \chi^2(|\Delta x|) = \chi^2(|x-\overline{x}|) = \chi_\nu^2(x) + \chi_{\overline{\nu}}^2(\overline{x})\,,
\end{equation}
where $\chi_\nu^2$ $(\chi_{\overline{\nu}}^2)$ is the $\chi^2$-function for the analysis of neutrino (antineutrino) data. Note that this is not the standard procedure where one usually considers the same value for neutrino and antineutrino oscillation parameters ($x$ = $\overline{x}$). This approach, however, can lead to impostor solutions, as shown in Ref.~\cite{Barenboim:2017ewj}.

\begin{figure}
\centering
\includegraphics[width=0.48\textwidth]{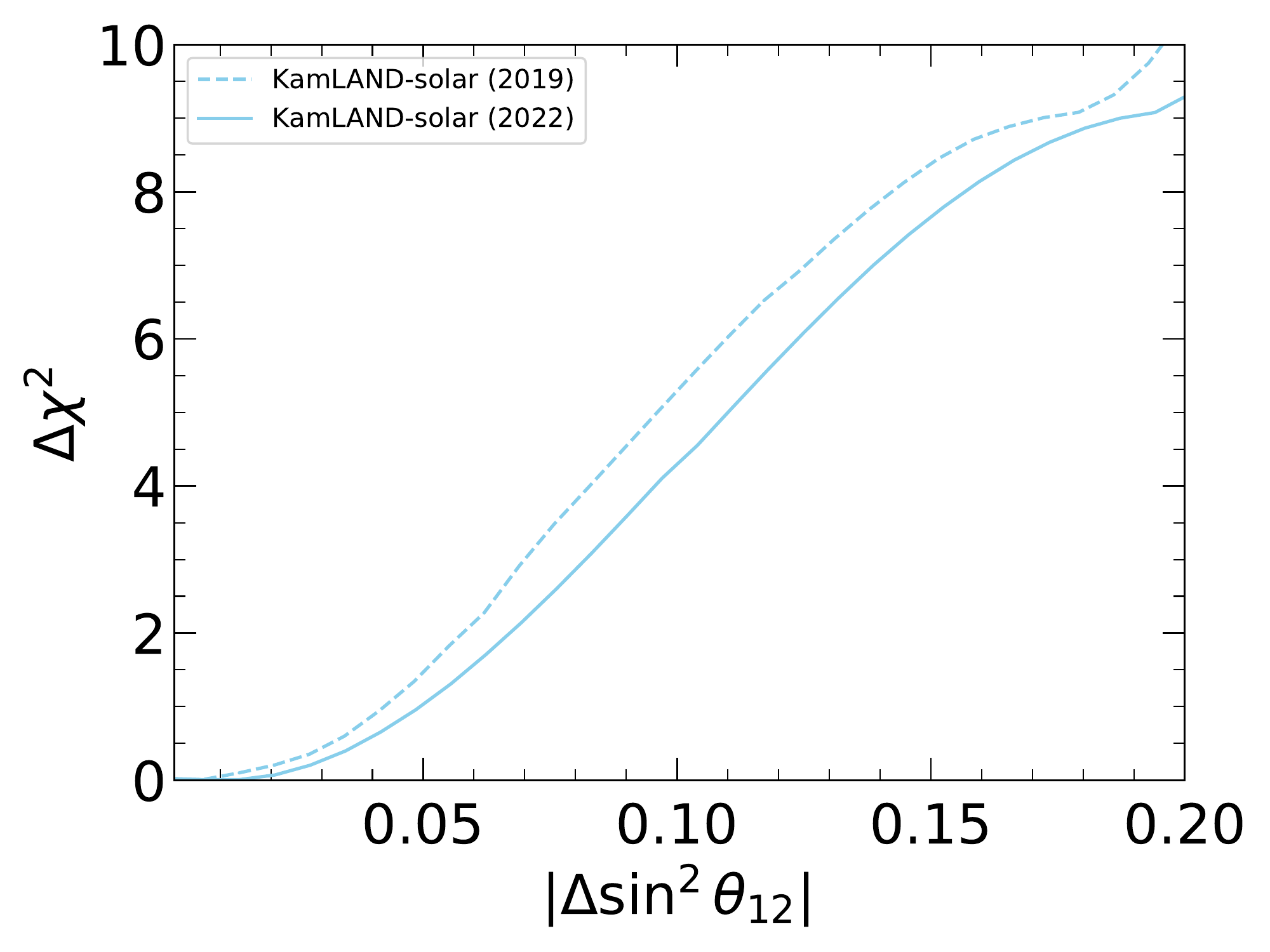}
\includegraphics[width=0.48\linewidth]{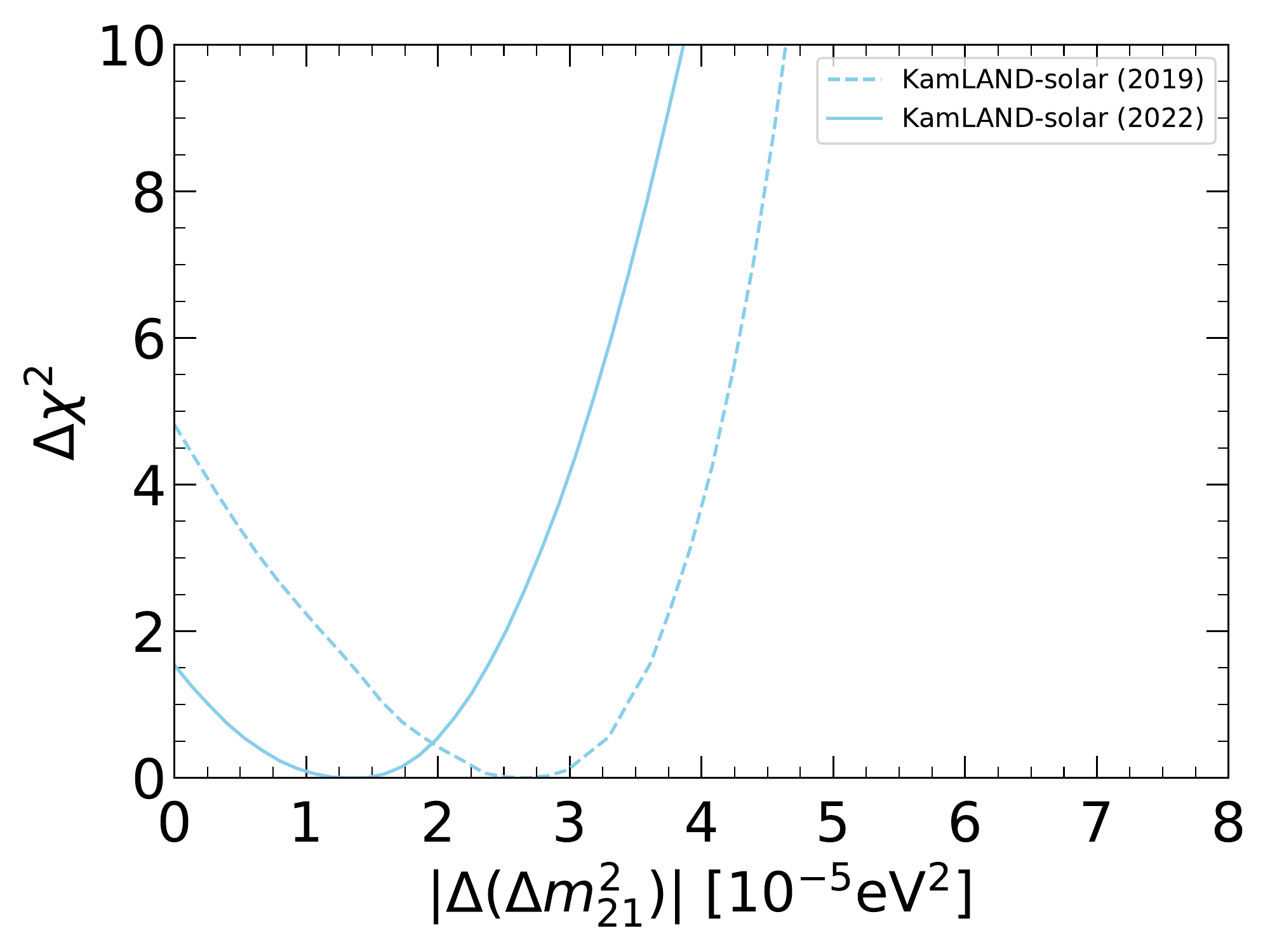}
\caption{\label{fig:CPT_curr}
Limits on CPT-violating neutrino oscillation parameters in the solar sector using KamLAND and solar neutrino data. In both panels, solid (dashed) lines correspond to the analysis with the 2022 (2019) Super-Kamiokande solar data set.}
\end{figure}

The constraints reported in Refs.~\cite{Barenboim:2017ewj,Tortola:2020ncu} relied on previous measurements of solar neutrinos~\cite{Cleveland:1998nv,Kaether:2010ag,Abdurashitov:2009tn,Bellini:2011rx,Bellini:2013lnn,Hosaka:2005um,Cravens:2008aa,Abe:2010hy,Nakano:PhD}. Here we update those limits using the most recent preliminary results from the Super-Kamiokande Collaboration \cite{yusuke_koshio_2022_6695966}. 
In Fig.~\ref{fig:CPT_curr} we compare the bound that is obtained using solar data as of 2022 and the one obtained from the older solar data set discussed in Ref.~\cite{deSalas:2020pgw}. 
From the left panel of Fig.~\ref{fig:CPT_curr}, one sees that the bound on $|\Delta\sin^2\theta_{12}|$ is a bit weaker now. This is due to the fact that the current best fit value for the solar mixing angle ($\sin^2\theta_{12}=0.305$) is in slightly less agreement with the KamLAND best fit point ($\sin^2\theta_{12}=0.316$) than the previous solar best fit value ($\sin^2\theta_{12}=0.320$). It is only a tiny difference, but it propagates to the derived limit on CPT violation.
The results for $|\Delta(\Delta m_{21}^2)|$ are shown in the right panel of Fig.~\ref{fig:CPT_curr}. 
In this case, the most striking feature is the better agreement between the solar and KamLAND measurements, compared to the results obtained in 2019. Now the significance of the tension  between  both determinations, i.e. the value of $\Delta \chi2$  for $|\Delta(\Delta m_{21}^2)|$= 0, is lowered from $\sim2.2\sigma$ to $\sim1.2\sigma$.
The new bounds on CPT violation in the neutrino sector at 3$\sigma$ are therefore
 \begin{eqnarray}
 |\Delta(\Delta m_{21}^2)| = & |\Delta m_{21}^2-\Delta \overline{m}_{21}^2| &< 3.7\times 10^{-5} ~\text{eV}^2, 
  \nonumber \\
   |\Delta(\Delta m_{31}^2)| = & |\Delta m_{31}^2-\Delta \overline{m}_{31}^2| &< 2.5\times 10^{-4} ~\text{eV}^2, 
 \nonumber \\
  |\Delta\sin^2\theta_{12}| = & |\sin^2\theta_{12}-\sin^2\overline{\theta}_{12}| &< 0.187,
  \\
  |\Delta\sin^2\theta_{13}| = & |\sin^2\theta_{13}-\sin^2\overline{\theta}_{13}| &< 0.029,
  \nonumber \\
   |\Delta\sin^2\theta_{23}| = & |\sin^2\theta_{23}-\sin^2\overline{\theta}_{23}| &< 0.19\nonumber ,
 \label{eq:new-bounds}
 \end{eqnarray} 
where the constraints not related to the solar sector are taken from Ref.~\cite{Tortola:2020ncu}. Note that the updates in the long-baseline sector, Refs.~\cite{NOvA:2021nfi,T2K:2023smv}, which appeared after Ref.~\cite{Tortola:2020ncu} are not expected to change these bounds significantly.   

\section{Next generation experiments}
\label{sec:sim}
Although the current bounds on neutrino properties we have presented above represent the world's best absolute bound on CPT violation\footnote{Relative precision bounds, as the bound from the kaon system~\cite{ParticleDataGroup:2022pth}, depend on the choice of the scale used in the denominator, the averaged mass of the kaons, which clearly is not associated to CPT in any way.}, next-generation experiments have the potential to improve them to an impressive level.
In this section, we discuss the simulation details of the next generation of most CPT-wise relevant experiments. Regarding antineutrinos, we consider JUNO, while for the neutrino analysis we focus on the solar analyses of Hyper-Kamiokande and DUNE. 

\subsection{Reactor antineutrinos at JUNO}
The Jiangmen Underground Neutrino Observatory (JUNO)~\cite{JUNO:2015zny} is a next-generation medium-baseline reactor experiment. Its main scientific goals include the precise measurement of the solar neutrino oscillation parameters, the atmospheric mass splitting, and the neutrino mass ordering, apart from many other physics opportunities.
JUNO will consist of eight reactors~\cite{IceCube-Gen2:2019fet,JUNO:2021vlw} located at around 52~km distance from the main detector.
The main signal at the JUNO detector will come from the two 4.6 GW reactor cores at the Taishan power plant and the six 2.9 GW reactors at the Yangjiang power plant.
Our simulation of the experiment follows the descriptions of Refs.~\cite{JUNO:2015zny,IceCube-Gen2:2019fet,JUNO:2021vlw}, and in particular the one of Ref.~\cite{Forero:2021lax}.
Regarding the reactor flux, we use the parameterization of Huber-Mueller ~\cite{Mueller:2011nm,Huber:2011wv}. The fission fractions are assumed to be the same for all reactors and are kept fixed at $f_{235} = 0.561$, $f_{238} = 0.076$, $f_{239} = 0.307$, $f_{241} = 0.056$~\cite{JUNO:2020ijm}.
We assume a detector with a fiducial mass of 20~kt, a  selection efficiency of 82.2\%~\cite{JUNO:2021vlw} and an  energy resolution of 2.9\%~\cite{JUNO:update}. The cross-section is taken from Ref.~\cite{Vogel:1999zy}. 
An important source of background in JUNO is the contribution of antineutrinos from the nuclear power plants at Daya Bay and Huizhou, which depends on the oscillation parameters. This signal has been included in our simulation as two independent reactors with 17.4~GW thermal power at 215 and 265~km distance, respectively. We also include accidental, fast neutron, $^9$Li/$^8$He, $\alpha-$n and geo-neutrino background components, as extracted from Refs.~\cite{JUNO:2015zny,JUNO:2021vlw}.
Our statistical analysis considers an overall 2\% flux uncertainty, an 0.8\% uncertainty on the power of each core, and a 1\% uncorrelated shape uncertainty. The other relevant oscillation parameters ($\Delta \overline{m}_{31}^2$ and $\sin^2\overline{\theta}_{13}$) have been marginalized over in the analysis, although their effect on the precision of the solar neutrino parameters is negligible.

\subsection{Solar neutrinos at Hyper-Kamiokande}

Hyper-Kamiokande~\cite{Hyper-Kamiokande:2018ofw} will be the successor of the water Cherenkov detector Super-Kamiokande.  With a fiducial volume 8.3 times larger than its predecessor and improved photomultiplier detectors, it will provide  great capabilities for neutrino detection.
Hyper-Kamokande's rich physics program will include the study of solar, atmospheric and long-baseline accelerator neutrinos produced at the Japan Proton Accelerator Research Complex (J-PARC).
The relevant process for the study of solar neutrinos is elastic scattering on electrons, \textit{i.e.} $\nu_\alpha + e^- \rightarrow \nu_\alpha + e^-$. Note that, although this channel is sensitive to the three neutrino flavors, the cross-section for electron neutrinos, $\nu_e$, is larger than the one for $\nu_\mu$ and $\nu_\tau$.
We perform two different sensitivity analyses. In the first one, our simulation follows the approach taken in Ref.~\cite{Martinez-Mirave:2021cvh}, where 
we assume the same energy resolution as in Super-Kamiokande IV~\cite{Super-Kamiokande:2016yck}. Likewise, we consider an energy threshold of 5 MeV\footnote{Although Super-Kamiokande IV reached an energy threshold of 3.5 MeV, here we take a more conservative approach. In any case, the impact of the threshold is actually small, given the large energy uncorrelated uncertainties in the lowest-energy bins.}.
We perform an analysis of the day and night spectrum, taking into account the contributions from $^8$B and $hep$ neutrinos. The analysis considers the ratio between the oscillated and non-oscillated solar neutrino flux, as in \cite{Nakano:PhD, Martinez-Mirave:2021cvh}. We include a 4\% and 30\% uncertainty in the normalization of the $^8$B and $hep$ flux, respectively.
Following \cite{Martinez-Mirave:2021cvh,Nakano:PhD}, we incorporate in the analysis the uncertainty in the energy scale and uncorrelated systematic uncertainties of the same order of magnitude as those in Super-Kamiokande IV \cite{Super-Kamiokande:2016yck}. We also assume ten years of data-taking.
We refer to the simulation following these considerations as conservative.
Our second analysis takes into account several improvements expected for Hyper-Kamiokande and hence, it will be referred to as optimal. In the first place, and according to Ref.~\cite{Nishimura:2020eyq}, we consider an improvement by a factor of two in the efficiency. Secondly, we assume an energy threshold of 4.5 MeV~\cite{Hyper-Kamiokande:2018ofw}. Finally, we consider that both the energy resolution and the uncorrelated systematic uncertainties are improved by a factor of two with respect to Super-Kamiokande IV, resulting from the improvement of the photomultipliers. The remaining aspects of the simulation are identical to the conservative scenario. These considerations seem rather reasonable in the light of the preliminary sensitivity shown in Ref.~\cite{Yano:2021usb}. 

\subsection{Solar neutrinos at DUNE}

The Deep Underground Neutrino Experiment (DUNE)~\cite{DUNE:2020lwj,DUNE:2020ypp,DUNE:2020mra,DUNE:2020txw} is a
next-generation multipurpose neutrino experiment whose capabilities include the measurement of MeV neutrinos, among which, are solar neutrinos. Using the charged-current reaction $\nu_e + {}^{40}\text{Ar} \to e^- + {}^{40}\text{K}$, it will be sensitive to electron neutrinos from the Sun, with energies above 9 MeV. A lower energy threshold could be possible if a significant background reduction is achieved. 
Our sensitivity analysis considers 10 years of data and the projected full size of DUNE's far detector, consisting of 40 kton of liquid argon. For the cross-section, we consider the baseline configuration implemented in \texttt{SNOwGLoBES}~\cite{snowglobes} and we consider an energy resolution $\sigma(E)/E = 0.2$~\cite{DUNE:2020ypp,Castiglioni:2020tsu}.
We also include backgrounds from $^{222}$Rn and neutron capture~\cite{Pershey2020} with a 10\% uncertainty each, and we take into account the efficiency linearly increasing from 30\% at 9 MeV to 60\% at 21 MeV~\cite{Ilic2020}. Likewise, the uncertainty in the flux normalization for $^8$B and $hep$ neutrinos is the same as in the analysis carried out for Hyper-Kamiokande. 
The assumptions in this scenario are rather conservative and it might be possible that the background contribution could be reduced \cite{Zhu:2018rwc,Capozzi:2018dat,Borkum:2023dsu}. Moreover, existing studies suggest that a much better detection efficiency can be achieved in the MeV range~\cite{Ankowski:2016lab,Moller:2018kpn,DUNE:2020zfm}.
We perform two analyses in this paper: the first one uses only the spectral information for two angular bins, labeled as day and night, and the conservative choice of backgrounds and efficiencies, while in the second one, we separate the angular information for the night events into 10 bins of the same width and, in addition, consider the more favorable backgrounds and efficiencies. In particular, we consider a reduction of one order of magnitude in the neutron background rate and perfect efficiency. The first case can be considered as the conservative scenario, while the latter is an optimal scenario. 

\section{Future sensitivity to CPT violation in the solar sector}
\label{sec:future}

In this section, we discuss how well future experiments can improve the bound on CPT violation in the solar sector. We will consider the combinations of JUNO with Hyper-Kamiokande, with DUNE, and with a combined analysis of Hyper-Kamiokande and DUNE. As explained above, for both solar experiments we consider a conservative and an optimal experimental configuration. When discussing the combined analysis we consider always either both conservative or both optimal configurations.

\begin{figure}
\centering
\includegraphics[width=0.48\linewidth]{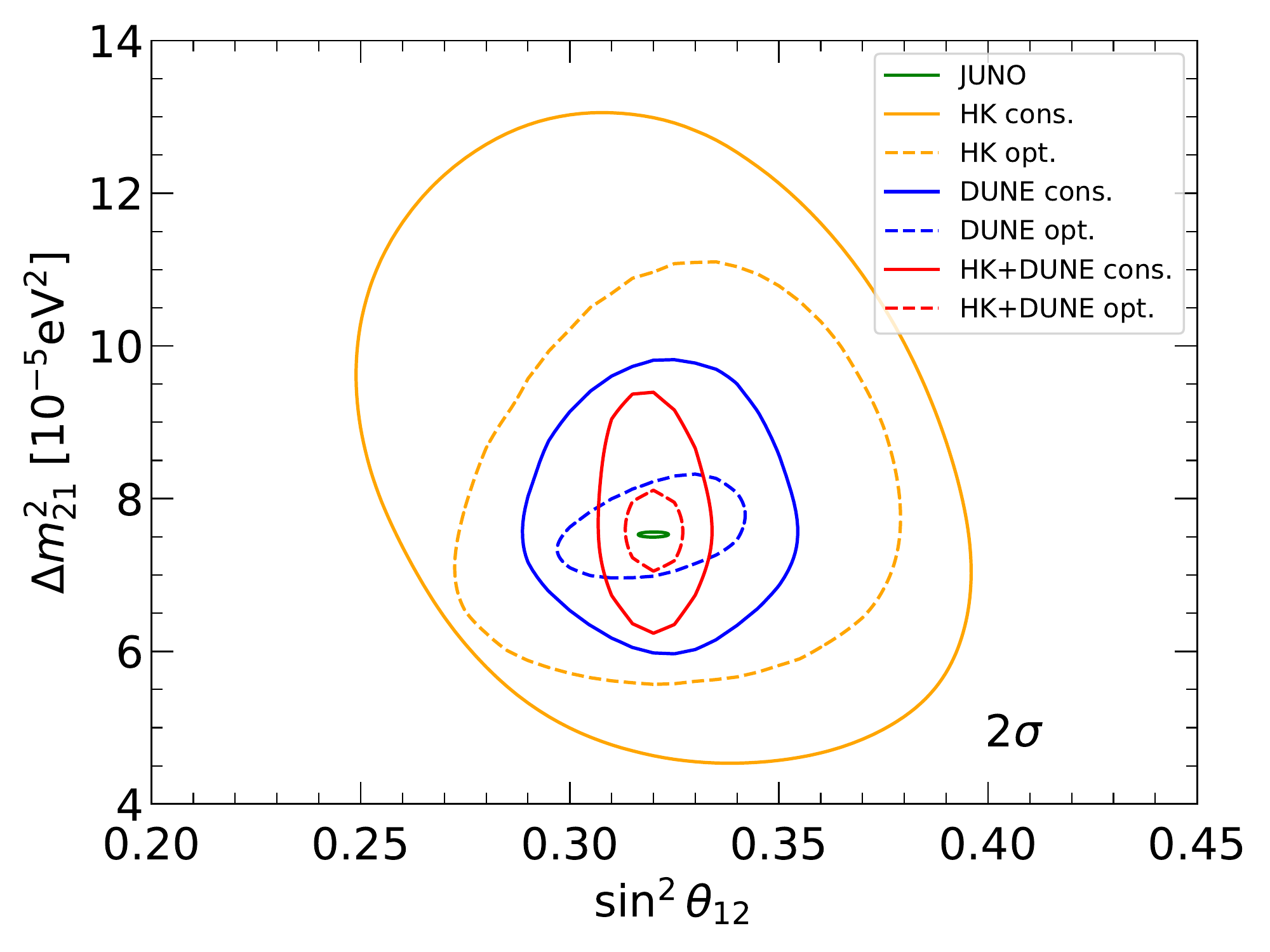}
\caption{\label{fig:solar_2D}
Expected 2$\sigma$ regions from Hyper-Kamiokande (HK), DUNE, HK+DUNE using conservative and optimal configurations, and JUNO, assuming the same oscillation parameters for neutrinos and antineutrinos.}
\end{figure}

In order to derive the projected sensitivities, we generate a CPT-conserving data set using $\sin^2\theta_{12}=\sin^2\overline\theta_{12}=0.32$ and $\Delta m_{21}^2=\Delta \overline m_{21}^2 = 7.53\times10^{-5}$~eV$^2$. 
The expected sensitivity of the different experiments to measure the solar parameters is shown in the left panel of Figure~\ref{fig:solar_2D}. 
It is clear that in the CPT-conserving scenario, the measurement will be dominated by JUNO, as indicated by the green contour.
The individual analyses of Hyper-Kamiokande and DUNE are shown in orange and blue (using the solid lines for the conservative cases and the dashed ones for the optimal cases), respectively. One can see that the solar sensitivity at DUNE is better than at Hyper-Kamiokande for the measurements of both solar oscillation parameters.
We obtain an interesting result for the combined analysis of both Hyper-Kamiokande and DUNE, see the red contours, where again the solid line is obtained from the conservative analyses, while the dashed lines are obtained for the optimal analyses. Note that the experiments share the systematic uncertainties related to the solar flux and, therefore, the combined sensitivity is much better than the simple sum of $\chi^2$ functions (which would be very similar to the DUNE-alone analysis). The reason for this improvement, particularly in the determination of the solar mixing angle, is that the use of two different detection channels allows breaking the degeneracy between the ${}^8$B flux normalization and $\sin^2\theta_{12}$.

\begin{figure}
\centering
\includegraphics[width=0.48\textwidth]{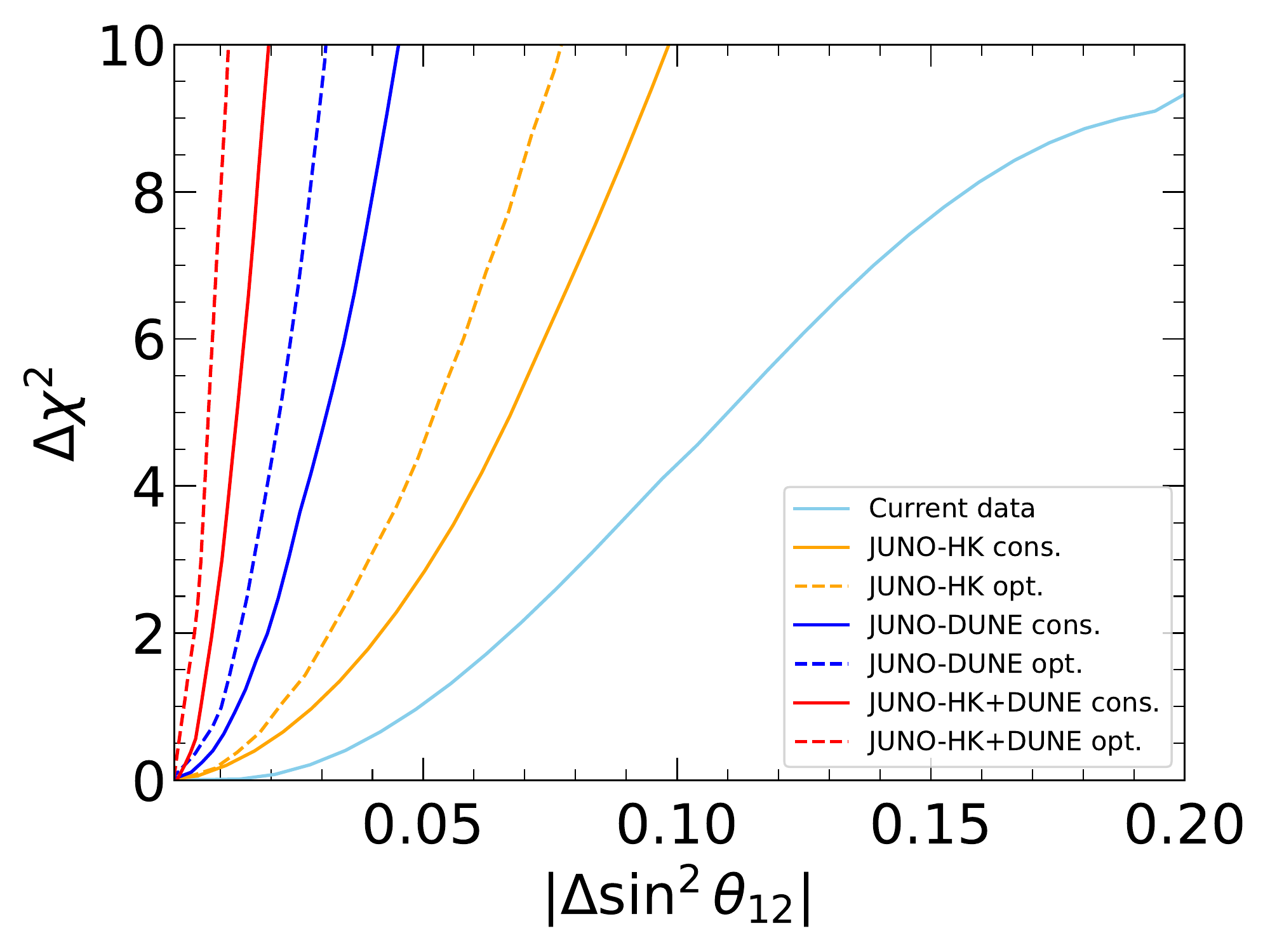}
\includegraphics[width=0.48\linewidth]{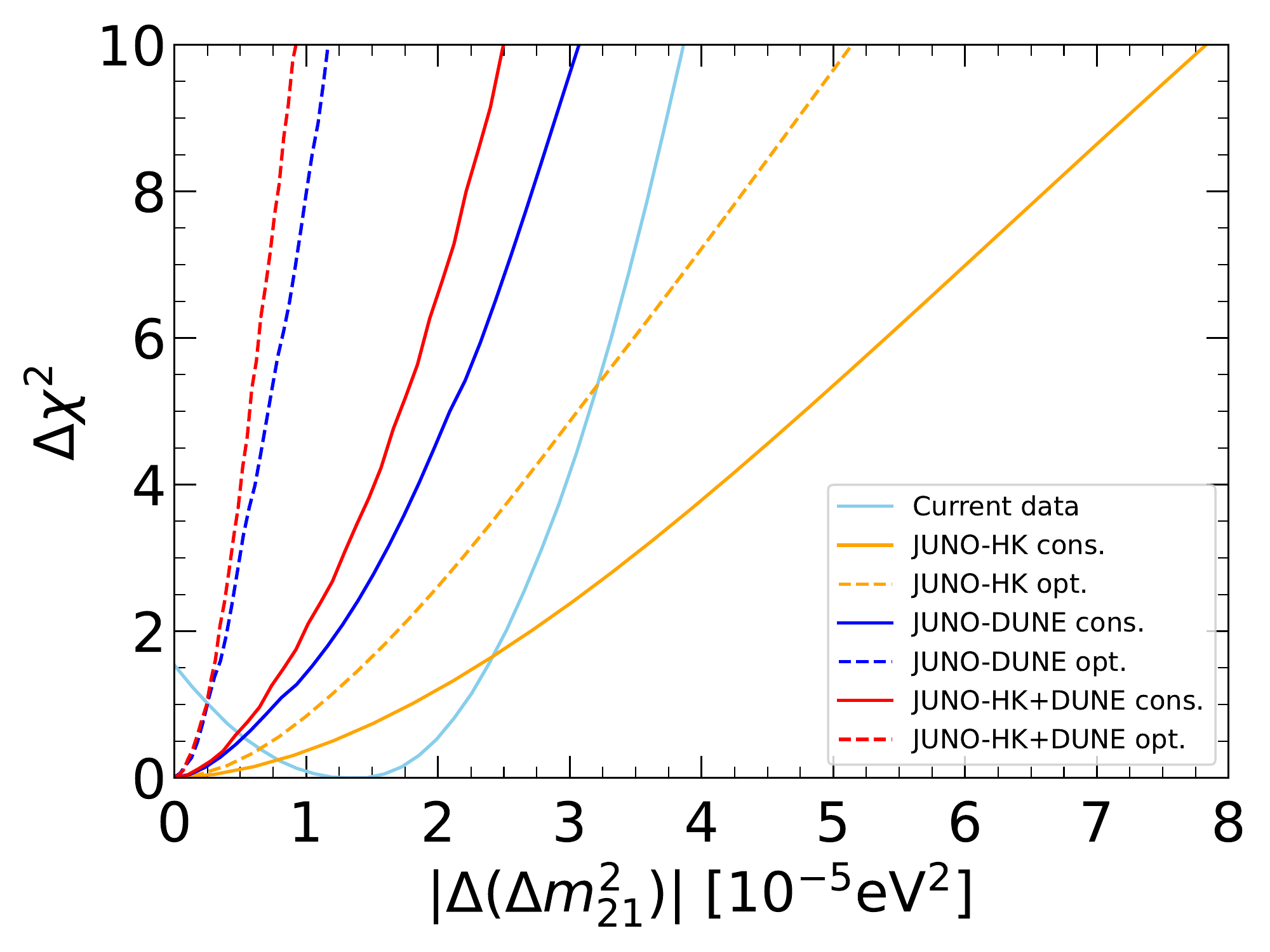}
\caption{\label{fig:CPT_sens}
Future sensitivity to CPT-violating neutrino and antineutrino oscillation parameters for different configurations and combinations of experiments in comparison with the current bounds.}
\end{figure}

Using the different solar analyses of neutrino data, and the antineutrino results from JUNO we can estimate the future sensitivity to CPT violation in the solar sector. The $\chi^2$ profiles for the differences of the neutrino and antineutrino oscillation parameters are shown in Fig.~\ref{fig:CPT_sens}. 
The left panel contains the sensitivity to constrain the difference in the solar mixing angle. Note that, even though the measurement of the mixing angle at Hyper-Kamiokande is rather weak, the potential limit from JUNO and Hyper-Kamiokande data will substantially improve the current 3$\sigma$ bound. This comes from JUNO's much more accurate determination of solar oscillation parameters compared to KamLAND.
As expected, the combination of JUNO and DUNE (see the solid and dashed blue lines in the left panel of Fig.~\ref{fig:CPT_sens}) shows a clear improvement in the sensitivity to $|\Delta\sin^2\theta_{12}|$. 
The joint analysis of JUNO, Hyper-Kamiokande and DUNE (red lines) is expected to improve the current limit by one order of magnitude.
Notice here that the impact of considering the conservative or the optimal configurations of DUNE and Hyper-Kamiokande in the combined analysis results in a difference of a factor two approximately in the limits on $|\Delta \sin^2\theta_{12}|$.
The expected $3\sigma$ sensitivities from the different analyses are summarized in the second column of Table~\ref{tab:cpt_sens}.

\begin{table}[b]
\centering
\begin{tabular}{|l||c|c|}
\hline
        & $~~|\Delta\sin^2\theta_{12}|~~$  &
      $|\Delta(\Delta m_{21}^2)|$ [$10^{-5}$eV$^2$] \\
      \hline
      current bound & \, 0.187  \,  & \,  3.7 \, \\
      \hline
        JUNO + HK conservative& \,  0.092 \, &  \,  7.2 \, \\
        \hline
        JUNO + HK optimal& \,  0.073 \, &  \,  4.7 \, \\
        \hline
        JUNO + DUNE conservative & \,  0.043 \,    & \,  2.9 \, \\
         \hline
        JUNO + DUNE optimal & \,   0.029 \, & \, 1.1 \, \\
        \hline
       JUNO + HK + DUNE conservative & \,   0.018 \, & \, 2.4 \, \\
        \hline 
    JUNO + HK + DUNE optimal &   \,   0.011 \,  &  \, 0.8 \,  \\
\hline
\end{tabular}
\caption{3$\sigma$ bounds on $|\Delta\sin^2\theta_{12}|$ and $|\Delta(\Delta m_{21}^2)|$ from current data in comparison with the future sensitivity expected from the combination of JUNO, Hyper-Kamiokande (HK) and DUNE for the conservative and optimal configurations.}
\label{tab:cpt_sens}
\end{table}

The sensitivity to $|\Delta(\Delta m_{21}^2)|$ is shown in the right panel of Figure~\ref{fig:CPT_sens}, and in the third column of Table~\ref{tab:cpt_sens}. 
As in the case of the mixing angles, the  sensitivity of the JUNO+Hyper-Kamiokande analysis is weaker than for JUNO+DUNE, and it is further improved when combining DUNE and Hyper-Kamiokande. 
Note, however, that here the improvement is not as strong as in the last case, since, as it was shown in Fig.~\ref{fig:solar_2D}, the combination of solar experiments is more effective for the determination of the mixing angle. 
On the other hand, in this case, the assumed configuration of DUNE has a strong impact on the result, as can be seen by comparing the blue and red solid lines with the dashed ones. Indeed, for the optimal setup of DUNE, the sensitivity to  $|\Delta(\Delta m_{21}^2)|$ is improved by more than a factor of two.

Summarizing, we find that the next generation of solar neutrino experiments in combination with the reactor experiment JUNO will be able to improve the current bounds on CPT-violating oscillation parameters for neutrinos and antineutrinos significantly, even by more than one order of magnitude in the case of the mixing angles.

\section{Measuring CPT violation}
\label{sec:cpt_measure}

\begin{figure}
\centering
\includegraphics[width=0.48\linewidth]{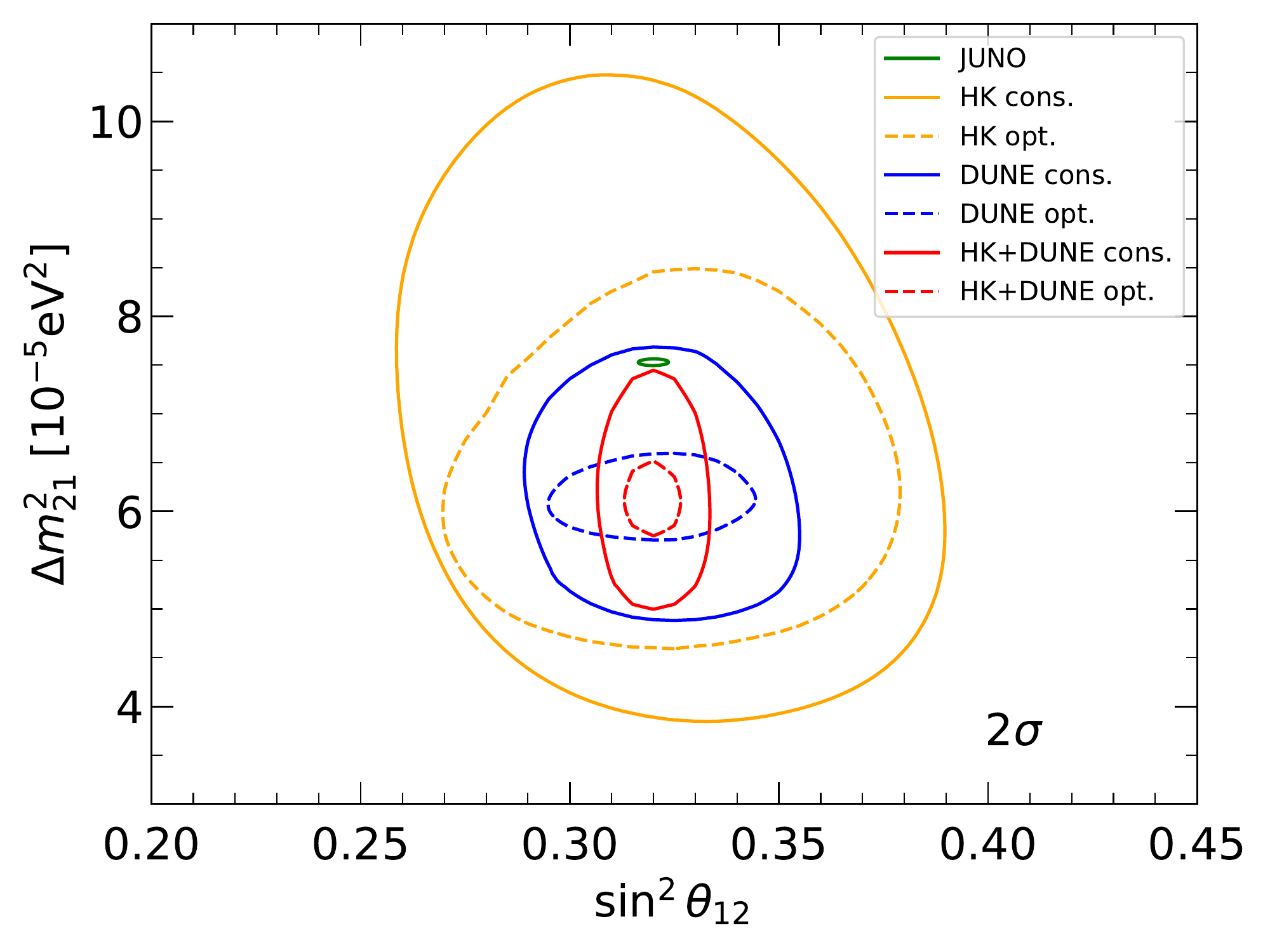}
\includegraphics[width=0.48\textwidth]{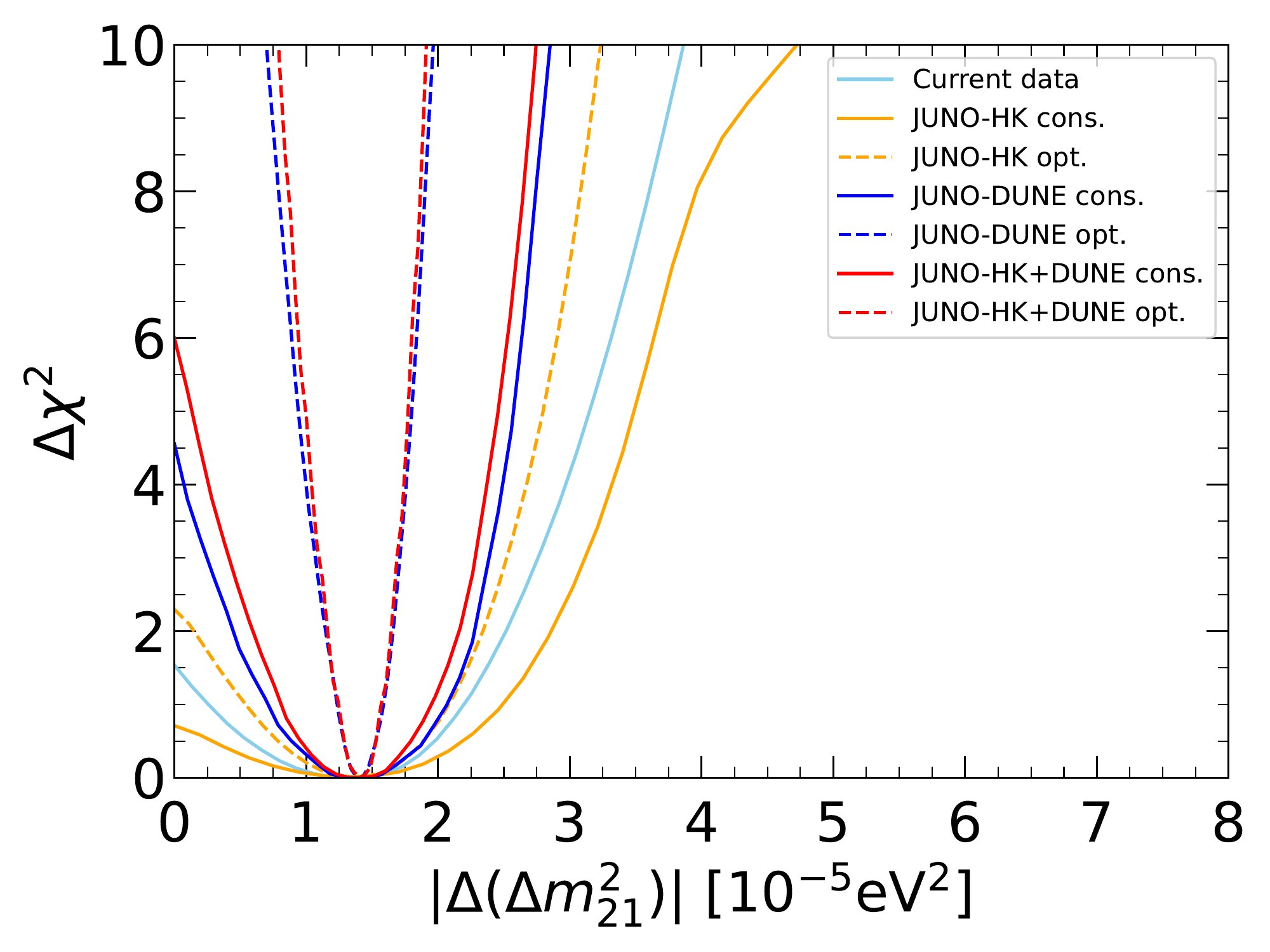}
\caption{\label{fig:CPT_sens_CPT}
Left: Expected $2\sigma$ regions from Hyper-Kamiokande (HK), DUNE, HK+DUNE using conservative and optimal configurations, and JUNO, assuming different oscillation parameters for neutrinos and antineutrinos. Right:  $\Delta\chi^2$ profiles of the difference of mass splittings, quantifying the tension for the different combinations.}
\end{figure}

In this section, we assume that CPT is violated in Nature. Then, we generate the antineutrino oscillation data using $\Delta \overline m_{21}^2 = 7.53\times10^{-5}$~eV$^2$, whereas neutrino oscillation data are generated using $\Delta m_{21}^2 = 6.10\times10^{-5}$~eV$^2$. 
We still assume $\sin^2\theta_{12}=\sin^2\overline\theta_{12}=0.32$, since the very small differences in the best fit values of the measurements of the mixing angle would most likely not be visible, not even with the next generation of neutrino experiments.
In the left panel of Figure~\ref{fig:CPT_sens_CPT} we show the expected region in the solar plane obtained from the different parameter settings. Note that the overall sensitivity of the solar experiments improves for smaller values of $\Delta m_{21}^2$, as can be seen by comparing the regions of Figure~\ref{fig:CPT_sens_CPT} with those of Figure~\ref{fig:solar_2D}. 
We observe the same behavior as before, namely the combination of DUNE and Hyper-Kamiokande improves mildly (significantly) the determination of the mass splitting (mixing angle) when comparing with the analysis of DUNE alone. Likewise,  the optimal DUNE configuration has a significant impact on the sensitivity to the solar mass splitting $\Delta m^2_{21}$. 

In the right panel of Figure~\ref{fig:CPT_sens_CPT} we show the $\Delta\chi^2$ profiles for the CPT-violating scenario. This quantifies how the current mild tension in the measurements of KamLAND and solar data could evolve in the future, in case the best-fit values of the measurements do not change significantly in either sector. 
Note that the tension from the combination of JUNO and Hyper-Kamiokande would be similar to the current one due essentially to KamLAND and Super-Kamiokande.
However, once we consider DUNE (DUNE+Hyper-Kamiokande), the tension is pushed to 2.0$\sigma$ (2.3$\sigma$) for the conservative configuration of both DUNE and Hyper-Kamiokande. The huge improvement of sensitivity seen in the left panel of Figure~\ref{fig:CPT_sens_CPT} when using the optimal configuration is also evident in the right panel. Actually, DUNE's optimal configuration could push the significance of the tension to 5.1$\sigma$ (5.9$\sigma$) for the analysis of DUNE alone (DUNE+Hyper-Kamiokande).

\section{Disentangling CPT violation and NSI}
\label{sec:cpt-nsi}

In this section we discuss the possibility of confusing CPT violation with neutrino nonstandard interactions. In order to do so we generate a CPT violating data set but analyze it in a CPT-conserving way including NSI. The same approach was followed in Ref.~\cite{Barenboim:2018lpo} in the context of accelerator neutrinos. Since the solar parameters are by far best measured by JUNO (and since JUNO is not as much affected by NSI), this reduces to generating a solar data set considering only solar neutrino oscillations with $\Delta m_{21}^2 = 6.10\times10^{-5}$~eV$^2$ and then trying to reconstruct it assuming $\Delta m_{21}^2 = 7.53\times10^{-5}$~eV$^2$ (the value measured by KamLAND) together with NSI. 

The evolution of neutrinos in the Sun can be described in an effective two-neutrino approach, as long as NSI are smaller or of the same order as neutrino interactions in the Standard Model. In that case, the survival probability is given by \cite{osti_6347879,Kuo:1989qe}
\begin{align}
    P_{ee}= \cos^4\theta_{13}P^{2\nu}_{ee} + \sin^4\theta_{13}\, ,
\end{align}
where $P^{2\nu}_{ee}$ is the electron neutrino survival probability derived in the effective two-neutrino framework from the Hamiltonian
\begin{align}
    H^{2\nu} = \frac{\Delta m^2_{21}}{4E} \begin{pmatrix}- \cos 2\theta_{12} & \sin2\theta_{12} \\ \sin 2\theta_{12} & \cos 2\theta_{12}\end{pmatrix} + \sqrt{2}G_F \Bigg[ \cos^2\theta_{13}N_e \begin{pmatrix}1& 0 \\ 0 & 0\end{pmatrix}+ \sum_{f = e, u, d} N_f \begin{pmatrix} 0 & \varepsilon_f \\ \varepsilon^*_f & \varepsilon'_f\end{pmatrix}\Bigg]\, .
    \label{eqn:2nu}
\end{align}
This effective Hamiltonian describes the evolution of the states $\nu$ = $(\nu_e,\, \nu_x)$, where $\nu_x$ is an admixture of the $\nu_\mu$ and $ \nu_\tau$ states. It depends on the number density of matter fields, $N_f$, with which neutrinos interact while propagating, \textit{i.e.} electrons, $u$-quarks and $d$-quarks. We have introduced the effective NSI parameters $\varepsilon_f$ and $\varepsilon'_f$, which are related to the NSI couplings in the neutral-current NSI Lagrangian \cite{Miranda:2004nb,Escrihuela:2009up,Friedland:2004pp,Esteban:2018ppq,Martinez-Mirave:2021cvh}. For simplicity, we only allow for NSI between neutrinos and $d$-quarks, denoting the corresponding NSI parameters, $\varepsilon_d$ and $\varepsilon'_d$ as $\varepsilon$ and $\varepsilon'$.

\begin{figure}
\centering
\includegraphics[width=0.48\linewidth]{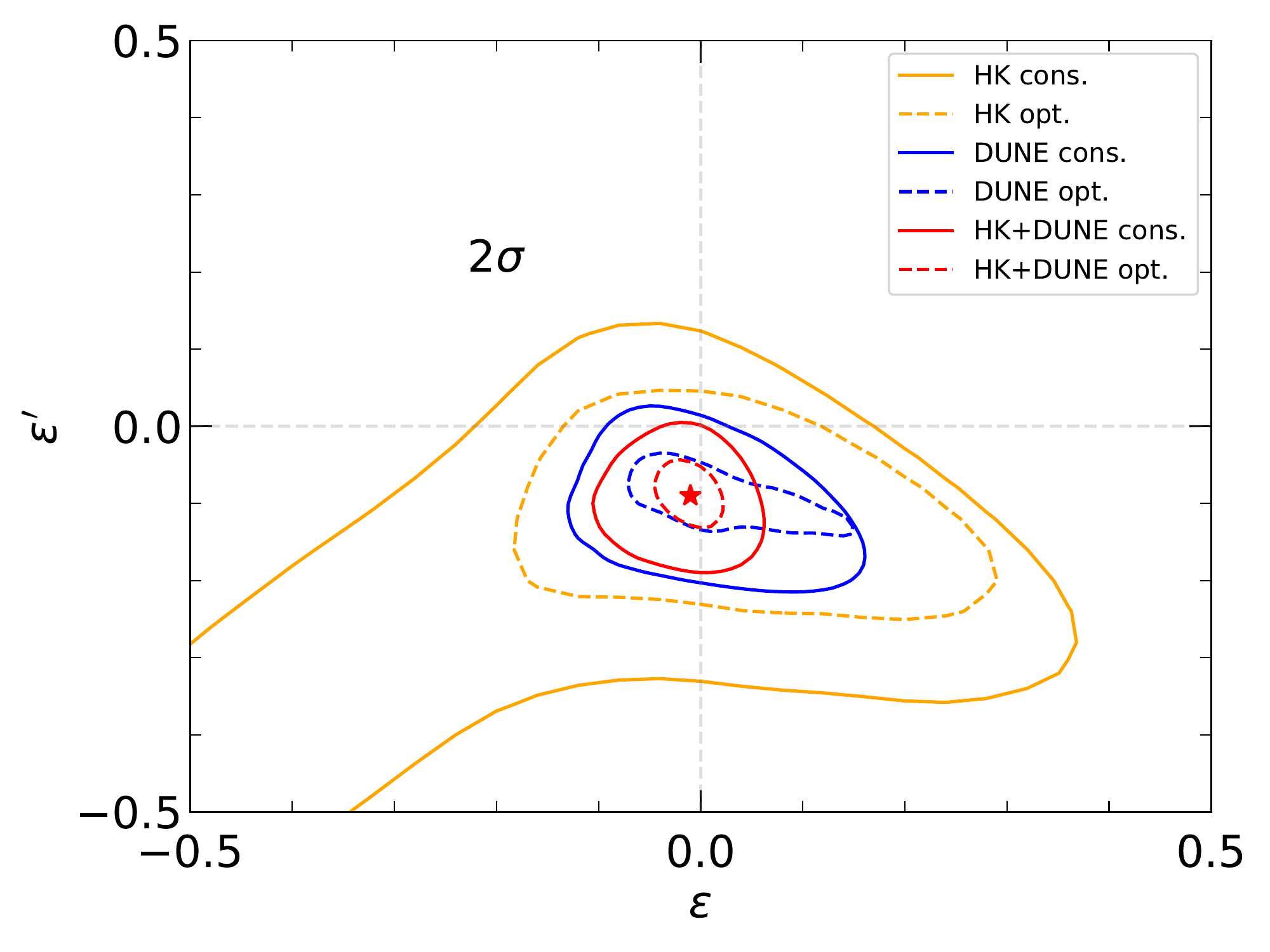}
\caption{
\label{fig:solar_nsi}
Regions in the NSI plane ($\varepsilon$, $\varepsilon'$) that can mimic the signal of CPT-violating solar neutrino mass splittings at the $2\sigma$ level in different experiments using conservative and optimal configurations.}
\end{figure}

The results from this analysis are shown in Fig.~\ref{fig:solar_nsi}. 
There, we plot the regions in the plane of the NSI parameters ($\varepsilon$, $\varepsilon'$) that can mimic the effect of CPT-violating neutrino mass splittings in the solar neutrino signal.
It should be noted that, for the separate analyses of Hyper-Kamiokande and the conservative setup of DUNE, the point $\varepsilon=\varepsilon'=0$ is within the 2$\sigma$ region. This is due to the fact that the tension between the two mass splittings, corresponding to $|\Delta (\Delta m^2_{21})| \simeq 1.4\times 10^{-5} \mathrm{eV}$, is quite small for these three data sets, as shown in the right panel of Fig.~\ref{fig:CPT_sens_CPT}. 
Combining the conservative analyses of DUNE and Hyper-Kamiokande restricts a bit more the allowed region, but $\varepsilon=\varepsilon'=0$ still lies at the border of the $2\sigma$ contour. 
Only when assuming the optimal DUNE configuration the standard scenario can be excluded at large confidence level as shown by the dashed contours in Fig.~\ref{fig:solar_nsi}.
However, even though the $\varepsilon=\varepsilon'=0$ value would be excluded from the analysis, the CPT-violating scenario could still be fit very well with NSI. Indeed, using the combination of optimal Hyper-Kamiokande and optimal DUNE the best fit value is found at $\varepsilon= -0.01$ and $\varepsilon'=-0.09$ and $\chi^2\approx4.7$. We therefore conclude  that this scenario of CPT violation could not be distinguished from NSI, unless independent constraints on $\varepsilon'$ from another type of experiment rule out the values preferred here.

\section{Conclusions}
\label{sec:conc}
The Standard Model of particle physics describes to an amazing precision all the experimental results obtained so far, with the exception of the neutrino sector. Neutrino masses were absent in the original model, and its incorporation necessarily brings along new physics. Not surprisingly, neutrinos themselves offer the opportunity not only to test the model itself, but also to test the paradigm where the model is immersed: local relativistic quantum field theory.

If  Nature can be described by local relativistic quantum field theory, CPT conservation is one of the few solid predictions of the paradigm,
and then particles and antiparticles are bounded to have the same mass and, if unstable, the same lifetime.
It is therefore extremely important to know whether the language we use to describe Nature is the correct one, and test it.

We do not know the origin of the neutrino mass, let alone its scale. However, we have measured neutrino mass squared differences to an admirable precision. In fact, neutrino mass squared differences are better known than some of the absolute masses of charged fermions and offer us a unique opportunity to test CPT conservation in an elementary particle system without charge contamination to an unprecedented level.  If CPT violation is related or originated from quantum gravity, and therefore suppressed by the Planck scale, naively we would expect its magnitude to be given by $ \langle v^2 \rangle /M_{\mbox{Planck}} \sim 10^{-5}$~eV, exactly in the ballpark the next generation of experiments will explore.

In this work, we have shown that CPT violation can be bounded or discovered by comparing antineutrino measurements at the reactor experiment JUNO with solar neutrino observations in DUNE and Hyper-Kamiokande. It is important to stress that DUNE improvements regarding angular bining and efficiencies have promoted the experiment to be a mayor player in the solar neutrino arena, as our results for the optimal setup of DUNE show. Improvements in energy resolution and a reduction of the energy-uncorrelated systematics would also improve significantly the solar neutrino results of Hyper-Kamiokande.

Had a positive CPT-violating signal been found, we have explored if it could be explained with NSI. Our results show that both sources of new physics can actually be confused in solar neutrino experiments. This highlights the relevance of combining different types of experiments, such as solar and neutrino scattering experiments to search for physics beyond the Standard Model~\cite{Escrihuela:2009up,Coloma:2017egw}.


\section*{Acknowledgments}

This work has been supported by a STSM Grant from COST Action “Quantum gravity phenomenology in the multi-messenger approach”, CA18108.
 C.A.T. is thankful for the hospitality at IFIC, Valencia where this work has been initiated.
C.A.T. is supported by the research grant ``The Dark Universe: A Synergic Multimessenger Approach'' number 2017X7X85K under the program ``PRIN 2017'' funded by the Italian Ministero dell'Istruzione, Universit\`a e della Ricerca (MIUR), by a {\sl Departments of Excellence} grant awarded by MIUR and the research grant {\sl TAsP (Theoretical Astroparticle Physics)} funded by Istituto Nazionale di Fisica Nucleare (INFN).
P.M.M. is supported by the Spanish fellowship FPU18/04571.
This work has also been supported by the Spanish grants PID2020-113775GB-I00 (AEI/10.13039/501100011033), PID2020-113334GB-I00 / AEI / 10.13039/501100011033  and CIPROM/2021/054 (Generalitat Valenciana). 
G.B. has received support from the European Union’s Horizon 2020 research and innovation program under the Marie Sk\l{}odowska-Curie grant agreement 860881-HIDDeN.

\bibliographystyle{JHEP}
\bibliography{bibliography}  

\end{document}